\documentclass[aps,prb,altaffillsymbol,amsmath,amssymb,amsfonts,superscriptaddress,reprint,noeprint]{revtex4-1}
\usepackage{graphicx}
\usepackage{tabularx}
\usepackage{color}
\usepackage{bm}
\usepackage{latexsym}
\usepackage{dsfont}
\usepackage{epstopdf}
\usepackage{CJK}
\usepackage{gensymb}
\usepackage{multirow}
\usepackage{hyperref}

\usepackage[version=4]{mhchem}

\begin{document}


\title{Low-dimensional quantum magnetism in Cu(NCS)$_2$: A molecular framework material}

\author{Matthew J. Cliffe}
\affiliation{Department of Chemistry, Lensfield Road, University of Cambridge, CB2 1EW, United Kingdom}

\author{Jeongjae Lee}
\affiliation{Department of Chemistry, Lensfield Road, University of Cambridge, CB2 1EW, United Kingdom}

\author{Joseph A. M. Paddison}
\affiliation{Churchill College, University of Cambridge, Storey's Way, Cambridge CB3 0DS, United Kingdom}

\author{Sam Schott}
\affiliation{Cavendish Laboratory, Department of Physics, University of Cambridge, JJ Thomson Avenue, Cambridge CB3 0HE, United Kingdom}

\author{Paromita Mukherjee}
\affiliation{Cavendish Laboratory, Department of Physics, University of Cambridge, JJ Thomson Avenue, Cambridge CB3 0HE, United Kingdom}

\author{Michael W. Gaultois}
\affiliation{Department of Chemistry, Lensfield Road, University of Cambridge, CB2 1EW,  United Kingdom}

\author{Pascal Manuel}
\affiliation{ISIS Facility, STFC Rutherford Appleton Laboratory, Harwell Oxford, Didcot OX11 0QX, United Kingdom}

\author{Henning Sirringhaus}
\affiliation{Cavendish Laboratory, Department of Physics, University of Cambridge, JJ Thomson Avenue, Cambridge CB3 0HE, United Kingdom}

\author{Si\^{a}n E. Dutton}
\affiliation{Cavendish Laboratory, Department of Physics, University of Cambridge, JJ Thomson Avenue, Cambridge CB3 0HE, United Kingdom}

\author{Clare P. Grey}
\affiliation{Department of Chemistry, Lensfield Road, University of Cambridge, CB2 1EW,  United Kingdom}
\email[]{cpg27@cam.ac.uk}
\date{\today}

\begin{abstract}
Low-dimensional magnetic materials with spin-$\frac{1}{2}$ moments can host a range of exotic magnetic phenomena due to the intrinsic importance of quantum fluctuations to their behavior.
Here, we report the structure, magnetic structure and magnetic properties of copper(II) thiocyanate, \ce{Cu(NCS)2}, a one-dimensional coordination polymer which displays low-dimensional quantum magnetism. Magnetic susceptibility, electron paramagnetic resonance (EPR) spectroscopy, \ce{^{13}C} magic-angle spinning nuclear magnetic resonance (MASNMR) spectroscopy, and density functional theory (DFT) investigations indicate that \ce{Cu(NCS)2} behaves as a two-dimensional array of weakly coupled antiferromagnetic spin chains ($J_2 = 133(1)$\,K, $\alpha = J_1/J_2 = 0.08$). Powder neutron-diffraction measurements confirm that \ce{Cu(NCS)2} orders as a commensurate antiferromagnet below $T_\mathrm{N} = 12$\,K, with a strongly reduced ordered moment (0.3\,$\mu_\mathrm{B}$) due to quantum fluctuations.

\end{abstract}

\pacs{Valid PACS appear here}
\maketitle


\section{Introduction}

\begin{figure}
\centering
 \includegraphics{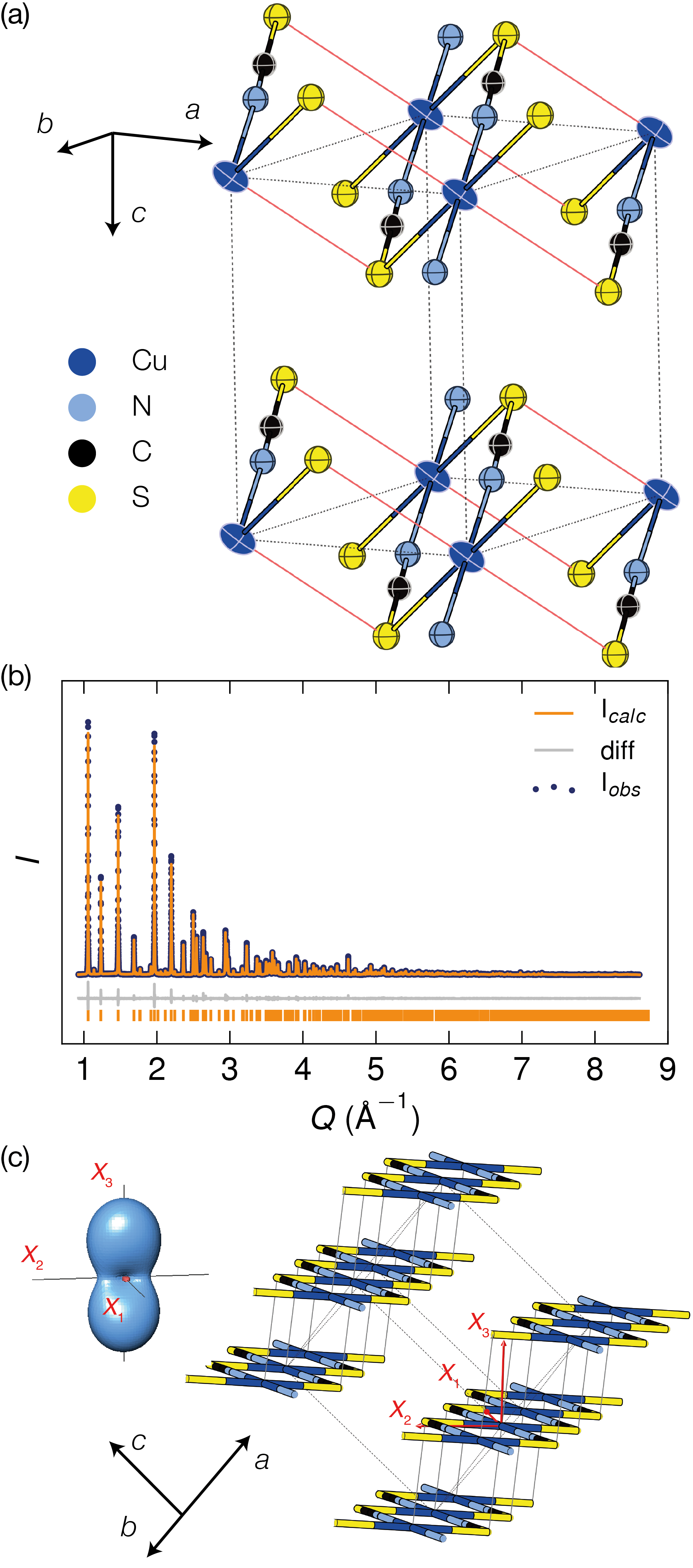}
 \caption{(a) Structure of \ce{Cu(NCS)2}. Copper atoms are shown in dark blue, sulfur in yellow, carbon in black and nitrogen in blue. The direction of the Jahn-Teller axis is indicated in red. Copper atoms were refined with anisotropic displacement parameters and the \ce{NCS-} ligand was refined isotropically. (b) Rietveld refinement of synchrotron powder X-ray diffraction data measured at 295\,K. (c) Orientation of the thermal expansivity tensor relative to the crystal structure of \ce{Cu(NCS)2}. Blue indicates positive thermal expansion, and red, negative. Color scheme for the structure as (a).}
 
 \label{fig:new_structure}
\end{figure}

Low-dimensional quantum magnets present some of the most interesting problems in condensed-matter physics, as they can represent theoretically-tractable examples of many-body quantum systems. One-dimensional quantum Heisenberg antiferromagnets (QHAFM) show a wide range of unusual physical phenomena, including multiferroicity,\cite{Zhao2012} fractionalized excitations\cite{Mourigal2013} and spin-Peierls transitions.\cite{Jacobs1976,Hase1993} Two-dimensional QHAFM materials are important model systems to explore the quantum fluctuations implicated in high-temperature superconductivity.\cite{Manousakis1991} Materials in which the magnetic behavior is between 1D and 2D limits, such as spin ladders and 2D lattices with spatially-anisotropic exchange interactions, therefore provide ideal opportunities to explore models of quantum magnetism.\cite{Dutton2012a,Dagotto1996,Keith2011}

Molecular frameworks---materials with framework structures in which at least one component is a molecule---are of current interest because they present a number of contrasts to conventional inorganic materials. The increased physical separation of magnetic centers allows for low-dimensional magnetic behavior,\cite{Liu2016c} while the large variety of molecular ligands permits fine-tuning of the magnetic interactions to a level beyond that achievable in inorganic frameworks.\cite{Yamada2017} Moreover, the use of larger molecular ligands produces open crystal structures, which can lead to permanent porosity allowing for guest-dependent magnetic behavior.\cite{Kepert2000} However, a limitation of molecular-framework materials has been that multi-atom ligands usually yield weak exchange interactions ($J\sim10$\,K) compared to inorganic materials. As a result, the systematic investigation of their magnetic properties is still relatively underdeveloped. Nevertheless, frameworks constructed using a number of simple ligands---including formate (\ce{HCOO^-}),\cite{Saines2015,Wang2013a,Canadillas-Delgado2012,DallaPiazza2014} pyrazine (\ce{NC4H4N}),\cite{Goddard2012,Manson2006} dicyanamide (\ce{NCNCN-})\cite{Batten2003,Lappas2003} and cyanide (\ce{CN-})\cite{Ferlay1995,Franz2004}---have been shown to exhibit a wide range of properties, including low-dimensional magnetism,\cite{Saines2015} multiferroicity,\cite{Canadillas-Delgado2012} fractionalisation\cite{DallaPiazza2014} and room-temperature ferrimagnetism.\cite{Mautner1996} 

In this paper, we study the molecular-framework material copper(II) thiocyanate, \ce{Cu(NCS)2}. This material is part of a series of binary transition metal thiocyanates that includes nickel(II),\cite{Dubler1982} cobalt(II),\cite{Shurdha2012} silver(I) \cite{Williams2007} and copper(I) thiocyanate.\cite{Kabesova1976} The existence of \ce{Cu(NCS)2} has been known for over a century,\cite{Philip1916,Hunter1969} but its structure has proved elusive due to its propensity to be spontaneously reduced to \ce{Cu(NCS)} in aqueous solution. Despite the instability of the parent compound, a number of complexes of copper(II) thiocyanate with co-ligands have been reported and their magnetic properties investigated.
\cite{Julve1993,Navarro1997,White1999,Shi2006,Machura2013,Wriedt2009} The magnetic interactions in these complexes are typically weak, as is usual for molecular-framework materials, due to the extended superexchange pathways through the auxiliary ligand.\cite{Bordallo2003} Even where Cu--NCS--Cu connectivity is present, the magnetic communication is often disrupted because the Jahn-Teller elongation of Cu$^{2+}$ tends to occur along the S--Cu--S direction, preventing efficient superexchange through the \ce{NCS-} ligand.\cite{Wriedt2009} Nevertheless, it has been shown that strong superexchange coupling through the  the \ce{NCS-} ligand is possible. \cite{Julve1993,Navarro1997,White1999,Shi2006} Moreover, the chemical similarities between thiocyanate and halide ligands suggest that copper(II) thiocyanate may have a structure analogous to its chloride and bromide equivalents, which are low-dimensional magnetic materials.\cite{Barraclough1964} These results suggest that the magnetic behavior of \ce{Cu(NCS)2} may be unconventional.

Here, we report the synthesis, crystal structure and magnetic properties of \ce{Cu(NCS)2}. Using powder X-ray and neutron diffraction measurements, we show that \ce{Cu(NCS)2} consists of Cu--NCS--Cu chains. Using a combination of magnetic susceptibility, electron paramagnetic resonance (EPR) and magic-angle spinning nuclear magnetic resonance (MASNMR) measurements, we show that its low-dimensional structure leads to low-dimensional magnetic behavior. We identify \ce{Cu(NCS)2} as a quasi-one-dimensional (1D) magnetic material with an intra-chain magnetic interaction $J_2 = 133\,$K that is anomalously strong for a molecular framework. This coupling is particularly notable as it occurs via a four-bond superexchange through the \ce{NCS-} ligand (Cu--N--C--S--Cu). We explain the mechanism of superexchange using quantum-chemical calculations. We use powder neutron diffraction to determine the ordered magnetic structure below $T_\mathrm{N} =12$\,K, which we find to be a G-type antiferromagnet with an ordered magnetic moment that is significantly reduced due to quantum fluctuations. 

\section{Methods}
\subsection{\label{sec:synthesis}Synthesis of {Cu(NCS)$_2$}}
\ce{Cu(NO3)2.2.5H2O} (2.33\,g, 10\,mmol) was dissolved in the minimum quantity of deionized \ce{H2O} (approx. 5\,ml), and rapidly added to a saturated solution of \ce{NH4NCS} (3.04\,g, 40\,mmol), giving an immediate black precipitate. This was stirred for one minute, before being filtered under vacuum and rapidly washed on the filter with 10\,ml \ce{H2O}. The residue of was then dried at 50\,$^\circ$C for one hour, giving a black microcrystalline powder (1.64\,g, 91\%). The product was analyzed for its elemental purity (combustion analysis for CHN and inductively-coupled plasma mass spectrometry for Cu and S). The measured (calculated) elemental composition was \ce{CuS2C2N2}: Cu 35.64\% (35.36\%), S 35.56\% (35.69\%), C 13.28\% (13.37\%), N 15.31 (15.59\%), H 0.0\% (0.0\%). This procedure, with all quantities scaled up by a factor of three, was used to synthesize the sample used for neutron-diffraction measurements.

\subsection{Synchrotron X-ray diffraction measurements}
A high-resolution synchrotron X-ray powder diffraction measurement of \ce{Cu(NCS)2} was carried out at beamline 11-BM at the Advanced Photon Source (APS) using a wavelength of 0.414537\,{\AA}. The sample was loaded into a 0.8\,mm diameter Kapton capillary. Rietveld refinement of the data was carried out using Topas Academic 4.1.\cite{Rietveld1969,Topas} Lattice parameters, atomic positions and displacement parameters (anisotropic for Cu, isotropic for \ce{NCS}) were allowed to refine freely, along with crystallographic size and strain (anisotropically modelled using fourth-order spherical harmonics). A minor (1.17(2) wt.\,\%) secondary phase of ${\alpha}$-\ce{Cu(NCS)} was also found to be present.

\subsection{Variable-temperature X-ray diffraction measurements}
Variable-temperature powder X-ray diffraction measurements of \ce{Cu(NCS)2} were carried out on a Bruker D8 laboratory diffractometer equipped with a Oxford Cryosystems PheniX cryostat using Cu-K${\alpha}$ radiation ($\lambda\,=\,1.540$\,{\AA}). Data sets were measured over the temperature range 13--250\,K and Pawley refinement of the data was carried out using Topas Academic 4.1\cite{Pawley1981,Topas} to extract the variation in lattice parameters. 

\subsection{Neutron-diffraction measurements}
Powder neutron-diffraction measurements were carried out on the WISH instrument at the ISIS pulsed neutron and muon source.\cite{Chapon2011} A 4.26\,g sample of \ce{Cu(NCS)2} was loaded into a thin-walled vanadium canister 9.9\,mm in diameter up to a height of 61\,mm to ensure the sample was fully illuminated by the beam (height 40\,mm). Diffraction measurements were carried out at 1.4\,K, 20\,K, 80\,K, 150\,K and 300\,K, with shorter measurements carried out in steps of 2\,K between 4\,K and 20\,K. The data were corrected for absorption using the Mantid software.\cite{Arnold2014}

\subsection{Neutron-diffraction analysis}
The nuclear structure determined from X-ray diffraction was confirmed by Rietveld refinement against neutron-diffraction data. Due to the weakness of the magnetic reflections, the magnetic structure was determined by refinement against data from which the nuclear Bragg peaks and background had been removed via subtraction of a high-temperature (averaged 16--20\,K) data set. The magnetic structure was determined by first indexing the magnetic Bragg peaks to determine the magnetic propagation vector, and then using symmetry-mode analysis to determine the allowed magnetic irreducible representations using the ISODISTORT software.\cite{Campbell2006} Keeping the scale factor determined from the nuclear Bragg Rietveld refinement fixed, we then refined the direction and magnitude of the ordered moment for the lowest-temperature (1.4\,K) data set using Topas Academic 6.0.\cite{Topas} The temperature dependence of the ordered moment over the range 4\,K to 14\,K was calculated by refining its magnitude while fixing its direction.

\subsection{EPR measurements}
EPR measurements were performed on a Bruker E500 X-band spectrometer with a ER 4122SHQE cavity at a microwave frequency of 9.385\,GHz. The external magnetic field was modulated at 100\,kHz with an amplitude of 0.2\,mT and the spectra were recorded as first harmonics. The microwave power was set to 0.02\,mW, sufficiently small to prevent any saturation of the resonance and concurrent line-width broadening. 
A sample of approximately 20\,mg was loaded in an EPR tube inside an Oxford Instruments ESR900 cryostat with a temperature stability better than 0.1\,K. Spectra were recorded in 1\,K steps for 5--15\,K, in 10\,K steps for 30--70\,K and in 20\,K steps for higher temperatures.
The EPR spectra were fitted to powder pattern line-shapes with anisotropic $g$ with the EasySpin Toolbox for MATLAB.\cite{Stoll2006}
The magnetic susceptibilities were calculated by double integration of the best-fit first-harmonic spectrum.

\subsection{Electronic-structure calculations}
Magnetic calculations were performed with the Vienna Ab-initio Simulation Package (VASP)\cite{kresse_ab_1993} employing spin-polarized Perdew-Burke-Ernzerhof (PBE) exchange-correlation functionals\cite{Perdew1996} with a plane-wave cutoff of 400\,eV. To correct for the self-interaction error, we used the rotationally invariant Hubbard-$U$ correction by Dudarev \textit{et al.},\cite{dudarev_electron-energy-loss_1998} with $U_\mathrm{eff}=U-J$ obtained through the linear-response method of Coccocioni \textit{et al}.\cite{cococcioni_linear_2005} A $2\times2\times2$-supercell containing 8 Cu atoms was used to determine the magnetic ground state and the magnetic exchange-coupling constants. Experimental cell parameters determined using synchrotron diffraction were used with $4\times4\times4$ Monkhorst-Pack reciprocal grid\cite{monkhorst_special_1976} and a $10^{-6}$\,eV self-consistent field (SCF) convergence limit.
		
Molecular calculations were performed in the Gaussian16 package\cite{frisch_gaussian16_2016} employing the 6-311G basis and spin-polarized Becke, three-parameter, Lee-Yang-Parr (B3LYP) functional. A cluster with 4 Cu atoms was used to reduce the effect of the ends of the chain. SCF cycles were converged to $10^{-6}$\,Hartree.
		
We carried out calculations using the CRYSTAL package\cite{dovesi_c_2014} to determine the hyperfine parameters. As the hyperfine interaction is inherently a core property, all-electron Gaussian basis sets, as employed by CRYSTAL, are expected to be more accurate. The molecular Gaussian basis sets of Sch{\"a}fer \textit{et al}. were used with diffuse functions removed.\cite{schafer_fully_1992} We made use of the B3LYP exchange-correlation functional with 20\% and 35\% of Hartree-Fock exchange (labelled Hyb20 and Hyb35, respectively), because this has been previously shown to yield good predictions of hyperfine parameters.\cite{kim_linking_2010,middlemiss_density_2013,lee_systematic_2017} Experimental cell parameters determined by synchrotron X-ray diffraction were used with a $4\times4\times4$ Monkhorst-Pack reciprocal grid and $10^{-7}$\,Hartree SCF convergence limit.

\subsection{NMR measurements}
NMR spectra were acquired on a 9.4\,T Bruker Avance II spectrometer operating at a Larmor frequency of 100.576\,MHz for \ce{^{13}C} with a conventional Bruker 4\,mm MAS low-$\gamma$ probe, with an MAS rate of 12.5\,kHz. The radio-frequency amplitude of 83\,kHz and \ce{^{13}C} shift (29.5\,ppm, lower peak) were calibrated on an adamantane reference. Spectra were acquired using a rotor-synchronized spin-echo pulse sequence with recycle delays of 0.1\,s. Variable temperature experiments were performed using a liquid nitrogen heat exchanger. Temperature calibration was performed using $^{207}$Pb NMR experiments on \ce{Pb(NO3)2}.\cite{Bielecki1995} Fitting of the spectra was performed assuming a Lorenztian lineshape using the Bruker Topspin 3.0 software.

\subsection{Physical property measurements}
Measurements of the magnetic susceptibility were carried out on a 20.5\,mg sample of \ce{Cu(NCS)2} using a Quantum Design Magnetic Property Measurement System 3 (MPMS) superconducting quantum interference device (SQUID) magnetometer. The zero-field cooled (ZFC) susceptibility was measured in an applied field of 0.01$\,$T over the temperature range $2$--$300\,$K. The small-field approximation for the susceptibility $\chi(T)\simeq\frac{M}{H}$, where $M$ is the magnetization and $H$ is the magnetic field intensity, was taken to be valid. Isothermal magnetization measurements over the field range $-4.5$ to $+4.5$\,T were carried out at 2\,K, 5\,K, 10\,K, 15\,K, 20\,K and 100\,K. Data were corrected for diamagnetism of the sample using Pascal's constants.\cite{Bain2008}

Heat-capacity measurements were carried out on a cold-pressed 13.1\,mg pellet of \ce{Cu(NCS)2} and silver powder (49 wt.\,\%), using a Quantum Design Dynacool Physical Property Measurement system (PPMS), between 2 and 300\,K. Contributions to the heat capacity due to silver were subtracted using tabulated values.\cite{Smith1995} Apiezon N grease was used to ensure good thermal contact. Data in the vicinity of the known thermal anomaly of the grease (265--285\,K) were neglected.\cite{Schnelle1999}

\section{Results and Discussion}
\subsection{Synthesis and structural characterisation}

\begin{table}
\caption{Atomic coordinates for \ce{Cu(NCS)2} determined by Rietveld refinement against synchrotron X-ray data at 295\,K. \label{tab_struct}} 
\begin{center}
\begin{tabular}{lll ll}
\hline \hline
\\
\multicolumn{2}{l}{Space group }&  \multicolumn{2}{l}{$P\overline{1}$}&\\

\\
$a$ (\AA) & 3.91596(2)& $\alpha$ ($^\circ$) &   82.37048(18)& \\
$b$ (\AA) & 5.65637(2)& $\beta$ ($^\circ$) & 85.07189(18)& \\
$c$ (\AA) & 6.06770(3)&  $\gamma$ ($^\circ$) & 113.49967(16)&\\
\\
	&$x$ 	& $y$ 	& $z$ 	& $B_{\mathrm{iso}}$ (\AA$^{2}$) \\ \hline
Cu & 0 & 0 & 0 & 2.04(4)* \\
C & $-0.2600(5)$ & $-0.4845(3)$ & $0.7875(3)$ & $1.30(4)$\\
N & $0.1288(4)$ & $-0.7154(3)$ & $0.1586(2)$ & $1.44(3)$\\
S & $-0.45039(14)$ & $0.22294(9)$ & $0.71543(8)$ & $1.611(14)$\\
\hline
 & $r$ ({\AA}) &  & $\theta$ ($^\circ$) &  \\
Cu--N & 1.9026(15) & Cu--N--S & 93.574(37)&\\
Cu--S & 2.41305(43) & N--C--S & 178.73(18) &\\
S--C & 1.6583(19) & Cu--N--C & 165.20(12) &\\
C--N & 1.1458(22) & Cu--S--C & 101.602(59) &\\
Cu--S$_\mathrm{JT}$ & 3.06613(57) & & &\\
\hline
\hline
\end{tabular}
Estimated standard errors are given in parentheses.
 *$B_{\mathrm{iso}}$ for Cu derived from anisotropic adps.
\end{center}
\end{table}

By adding a concentrated aqueous solution of copper(II) nitrate to a saturated aqueous solution of ammonium thiocyanate, and rapidly filtering and drying it, we were able to synthesize samples of \ce{Cu(NCS)2} as a microcrystalline powder with only minor impurities of Cu(NCS) (1.17(2) wt.\,\%). From laboratory powder X-ray diffraction measurements we determined that this material was identical to that previously reported.\cite{Hunter1969} We were able to index these data to a triclinic unit cell and then solve its structure using real-space methods and Rietveld refinement as implemented in Topas Academic 4.1.\cite{Topas} The crystal structure is shown in Fig.~\ref{fig:new_structure}(a). The Rietveld fit against synchrotron data collected at the 11-ID-B diffractometer at the APS is shown in Fig.~\ref{fig:new_structure}(b). 

The structure of \ce{Cu(NCS)2} can be conceptualized as a distorted variant of the layered \ce{M(NCS)2} structure (M=Co, Ni, Hg),\cite{Shurdha2012,Dubler1982,Beauchamp1972} where the orbital order of the Cu$^{2+}$ cations leads to an effective reduction in chemical dimensionality from a 2D sheet to a 1D chain [Fig.~\ref{fig:new_structure}(c)]. Despite the low symmetry of the structure, the presence of a center of symmetry at the copper enforces equivalence between all Cu--NCS--Cu interactions along the chain.
Variable-temperature powder X-ray diffraction measurements down to 13\,K confirmed the absence of any structural phase transitions. They also allowed us to establish the degree of thermal expansion in this material [Fig.~\ref{fig:new_structure}(c)]. As with many molecular framework materials, both the volumetric and linear thermal expansivities were significantly larger than for purely inorganic materials,\cite{Lappas2003} highlighting the increased relevance of phonons for their physical properties.\footnote{See Supplemental Material at XXX for additional characterisation data, including variable-temperature neutron and X-ray refinement data and diffuse reflectance spectra, further details on quantum-chemical calculations, and additional derivations.} Diffuse reflectance measurements on powder samples of \ce{Cu(NCS)2} showed strong absorbance across the visible region, with a band gap of 1.3(1)\,eV, likely corresponding to a strong ligand-to-metal charge transfer absorption.

\begin{figure}[h]
\centering
 \includegraphics[scale=0.5]{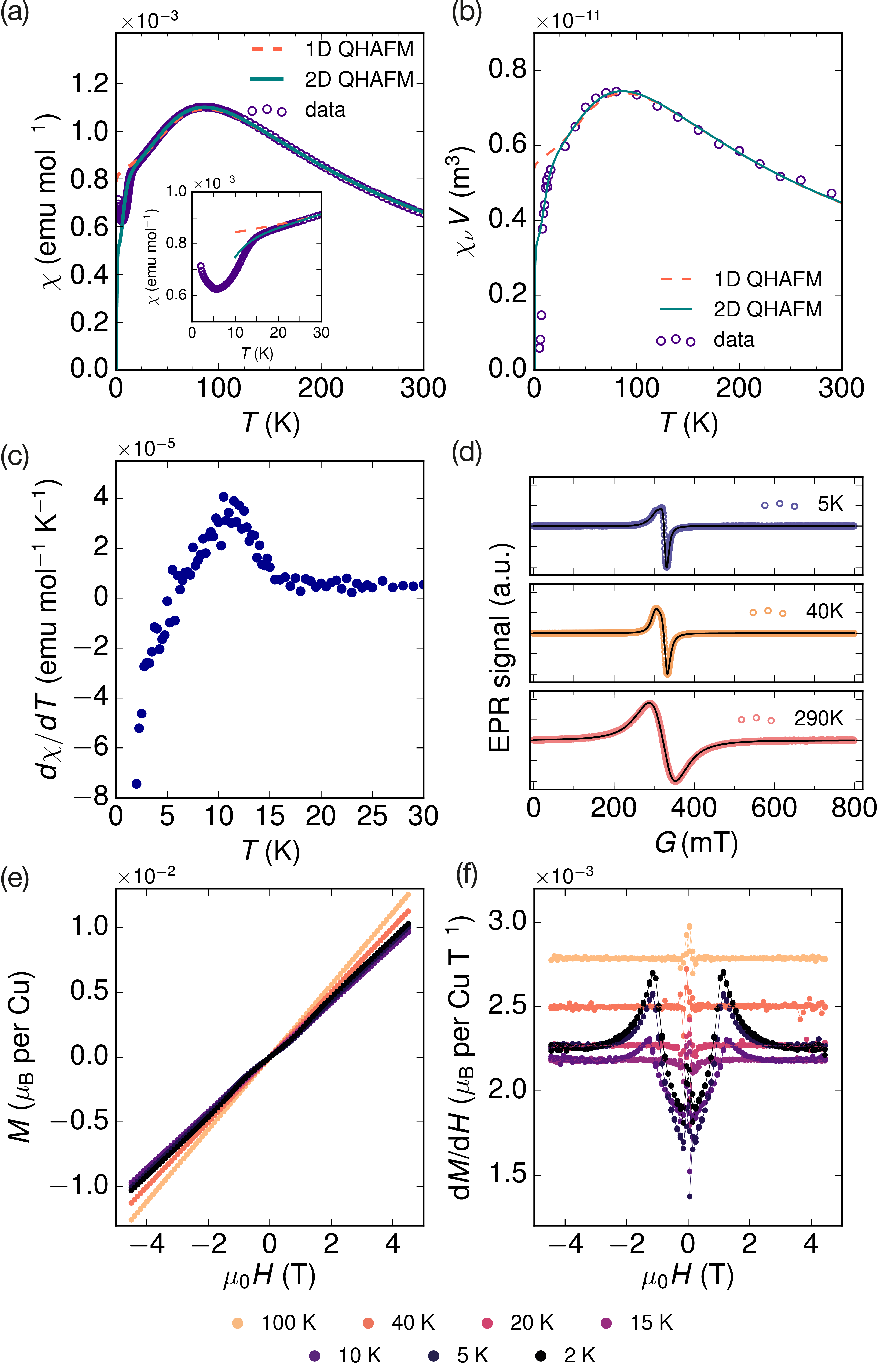}
 \caption{(a) Magnetic susceptibility measured using an MPMS in an applied field of 0.01\,T. Fits to 1D and 2D-QHAFM models over the temperature range 14--300\,K are shown as dashed and solid lines, respectively. (b) Magnetic susceptibility derived from integration of variable-temperature EPR spectra. Fits to 1D and 2D-QHAFM models over the range 14--300\,K are shown as dashed and solid lines, respectively. (c) Temperature derivative of the magnetic susceptibility. (d) Powder pattern fits to the EPR spectra. (e) Isothermal magnetisation measurements. (f) Field derivative of isothermal magnetisation measurements. For panels (e) and (f), temperature is indicated by the color (see key).}
 \label{fig:susceptibility}
\end{figure}

\subsection{Bulk magnetic measurements and single-ion properties}

The measured magnetic susceptibility [Fig.~\ref{fig:susceptibility}(a,b)] showed two principal features: a broad peak ($T_\mathrm{max}=86(2)$\,K) and a rapid decrease at lower temperatures ($T_\mathrm{N}=12.0$\,K). The anomaly at $T_\mathrm{N}$ is more clearly shown in the temperature derivative of the susceptibility [Fig.~\ref{fig:susceptibility}(c)]. The large ratio of these parameters $l=T_\mathrm{max}/T_\mathrm{N}=7.2$ suggests low-dimensional or frustrated magnetism.
\footnote{The conventional measure of frustration $f=T_\mathrm{CW}/T_\mathrm{N}$, where $T_\mathrm{CW}$  is the Curie-Weiss temperature, is difficult to apply to low dimensional materials, as the Curie-Weiss law is obeyed only at very high temperatures. For example, for a 1D QHAFM, to obtain a fitted value within 10\% of the true $T_\mathrm{CW}$ over the fitting range $T_\mathrm{fit}$ to $2T_\mathrm{fit}$, $T_\mathrm{fit}>8J$. For \ce{Cu(NCS)2}, this would require fitting above 1000\,K.\cite{Landee2014,Klumper2000}} 
At low temperatures ($<5$K) we observed a rise in susceptibility in our MPMS data [Fig 2(a)], which we ascribed to a small quantity of paramagnetic orphan spins (approximately 1.2\,\% via Curie law fitting). These orphan spins are commonly found in samples of low dimensional antiferromagnetic materials and are the result of small quantities of defects.\cite{Bono2005} We also carried out variable-temperature EPR measurements over the range 5--300\,K [Fig.~\ref{fig:susceptibility}(b)]. From these spectra, we determined the anisotropy of the $g$-tensor, $g_{z}=2.229$, $g_{y}=2.027$ and $g_{x}=2.047$ at 5\,K (temperature dependence is shown in Supplementary Information). The dominant contribution to the line-shape at low temperatures is the $g$-factor anisotropy; however additional homogeneous broadening from the finite spin lifetime and unresolved anisotropic broadening, likely from electronic dipole-dipole interactions, were necessary to account for the observed spectra [Fig.~\ref{fig:susceptibility}(d)]. The homogeneous broadening becomes more significant at higher temperatures, likely due to shorter spin-excitation lifetimes. The extracted susceptibility was consistent with data measured using the MPMS in the paramagnetic phase. Below $T_\mathrm{N}$, we found that the dynamic (measured by EPR) and static (measured by MPMS) susceptibilities began to diverge, with the dynamic susceptibility falling to zero. This divergence is typical upon ordering and has been observed in related low-dimensional magnetic materials.\cite{Sichelschmidt2002} We note, however, that the observed decrease in intensity, though rapid, occurs over a wider temperature range than ordinarily observed.

{\subsection{Spin Hamiltonian from quantum-chemical calculations}
\begin{figure}
\centering
 \includegraphics{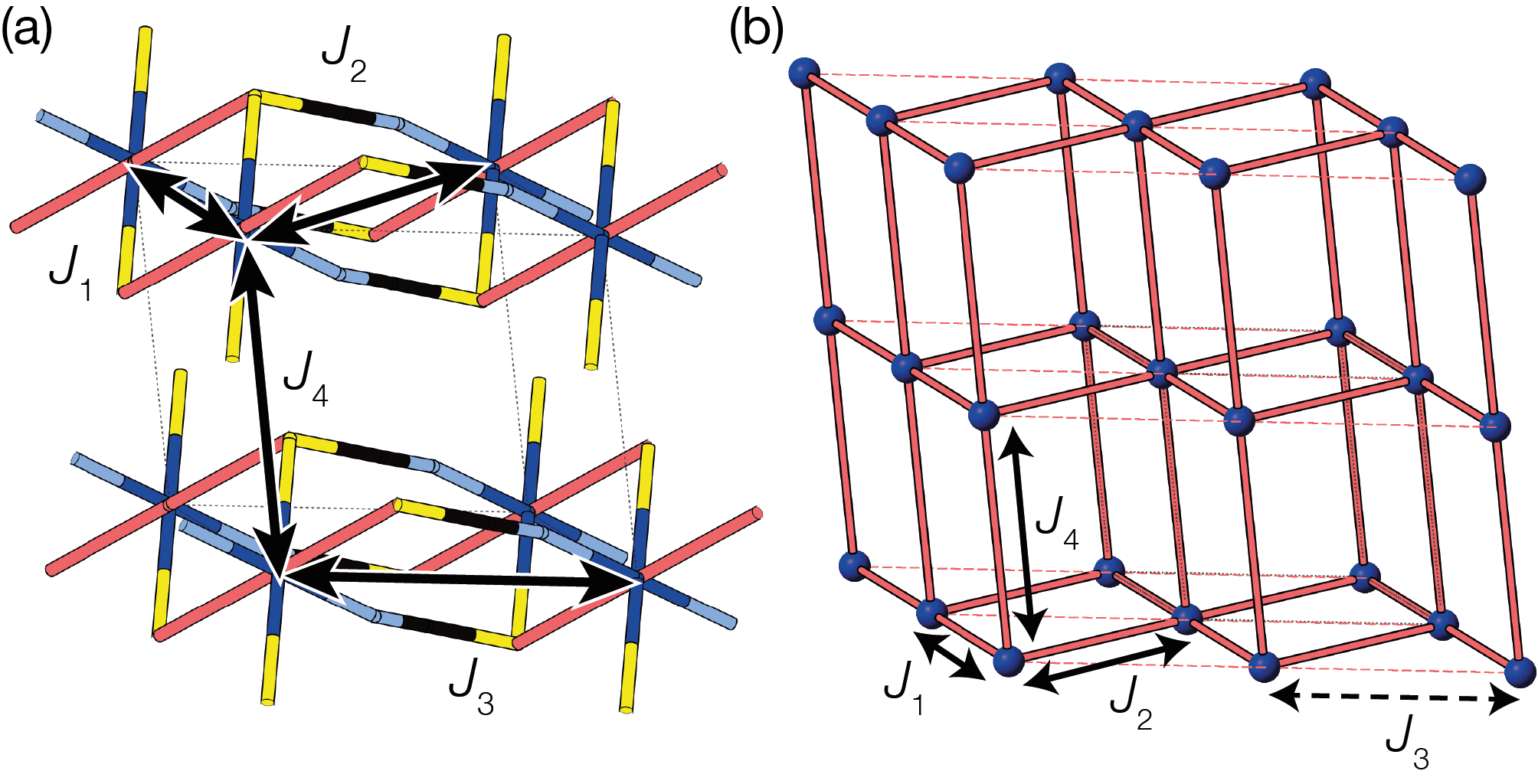}
 \caption{(a) The four nearest Cu--Cu distances and the four primary exchange interactions. (b) The magnetic lattice produced by considering these four interactions. As $J_3$ is small and does not cause any frustration it can be neglected, producing a simple cuboidal magnetic lattice with spatially anisotropic exchange.}
 \label{fig:mag_latt}
\end{figure}
The complexity of potential superexchange pathways typical of molecular frameworks led us to carry out quantum-chemical calculations to guide further analysis. We used spin-polarized density functional theory to calculate the magnetic superexchange interactions, predict the magnetic lattice, and propose a mechanism for the strong superexchange observed.
A linear-response calculation of the Hubbard-$U$ parameter for \ce{Cu(SCN)2} yields $U=7.35$\,eV, consistent with previous reports on copper oxide and sulfides.\cite{nolan_p-type_2006,suarez_structural_2016} Using this value for $U$, exchange interactions were calculated as follows: a $2\times2\times2$ supercell was constructed from the experimental structure, and then decorated with eight distinct magnetic orderings. The DFT+$U$ energies of these magnetic structures were used to perform a multidimensional linear regression to the Heisenberg exchange Hamiltonian
\begin{equation}
H = \sum_{i}\sum_{a}J_a\mathbf{S}_{i}\cdot\mathbf{S}_{i,a},
\end{equation}
where $J_a$ refers to the interaction between $a$-th closest Cu--Cu pair in the structure, $\mathbf{S}_i$ is the $i$-th spin, $\mathbf{S}_{i,a}$ is the spin connected to spin $\mathbf{S}_{i}$ by interaction $a$ and the Hamiltonian is defined for unique interactions (\emph{i.e.}, each pair is counted once). Exchange interactions are included for the four nearest Cu--Cu distances. We neglected magnetic dipolar coupling because the large Cu--Cu distances ($d_\mathrm{min} = 3.92\,${\AA) and small moment will give negligibly small dipolar interactions (on the order of 0.01\,K).
\begin{table}
\centering
\caption{\label{tab:J_values} DFT+$U$ derived interaction constants.}

\begin{tabular}{ l r r r r}
& $J_1$\,(K)& $J_2$\,(K)& $J_3$\,(K)& $J_4$\,(K)\\
$J_n$& $13.0(1.8)$ & $171.8(1.6)$& $-2.2(1.6)$& $1.6(0.8)$\\
\end{tabular}
\end{table}

\begin{figure}
	\centering
	\includegraphics{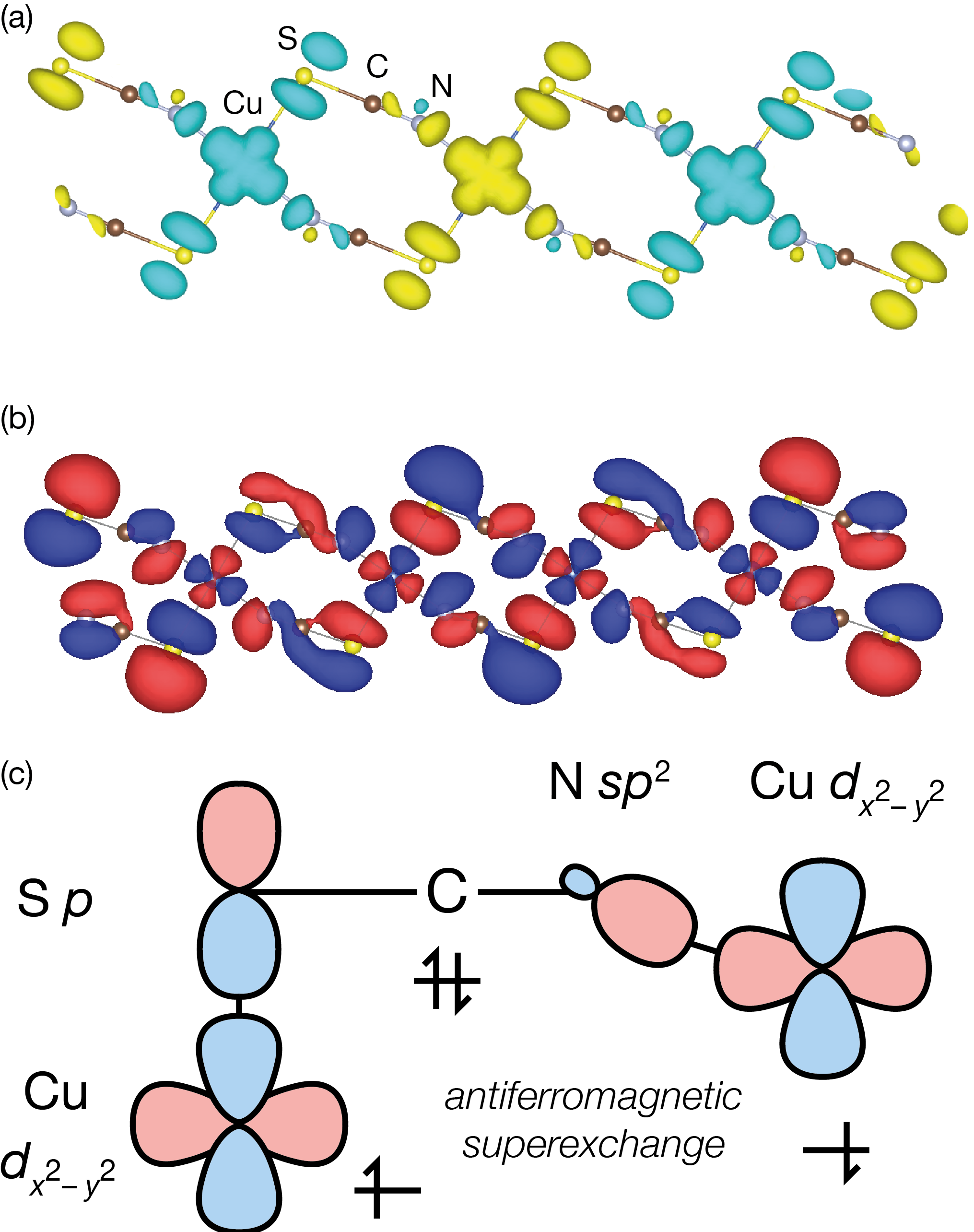}
	\caption{(a) Spin-density distribution in \ce{Cu(NCS)2}. Light blue and yellow colors indicate positive- and negative-spin moments, respectively. (b) The highest occupied molecular orbital in the \ce{Cu4} thiocycanate cluster. Phases are indicated by red and blue colors. (c) Diagram illustrating the orbitals involved in the superexchange interaction.}
	\label{fig:orbital}
\end{figure}

Using this approach, we obtained a self-consistent set of superexchange interactions $J_a$ for the four shortest Cu--Cu distances [Table~\ref{tab:J_values} and Fig.~\ref{fig:mag_latt}(a)], which in turn allowed us to predict the magnetic lattice of \ce{Cu(NCS)2} [Fig.~\ref{fig:mag_latt}(b)]. The lattice consists of antiferromagnetic chains ($J_2$) connected by order-of-magnitude weaker antiferromagnetic interactions ($J_1$)  forming a two-dimensional (2D) rectangular lattice. An even weaker diagonal ferromagnetic interaction $J_3$ is present within the layers, which does not frustrate $J_1$ and $J_2$ and so is unlikely to affect the magnetic behavior significantly. Finally, the layers are coupled antiferromagnetically by $J_4$, which is the shortest interlayer interaction [Fig.~\ref{fig:mag_latt}(a,b)]. Our calculations therefore predict that \ce{Cu(NCS)2} will behave as a 1D magnet at high temperatures, with weak interchain couplings becoming relevant at lower temperatures.

The strongest interaction ($J_2$) occurs along the Cu--N--C--S--chain, which can be understood by examining the  calculated spin density distribution [Fig.~\ref{fig:orbital}(a)]. Participation of the Cu $3d_{x^2-y^2}$ orbital is evident, as expected from the crystal-field model of square planar \ce{Cu^{2+}} ($d^9$). However, the most important observation is a strong spin delocalisation along the $\sigma$-bonds Cu--S and Cu--N. Antiferromagnetic intra-molecular coupling in the thiocyanate anion is observed between the two unpaired spins. To understand better the mechanism of this strong magnetic superexchange, we performed molecular calculations on a \ce{[Cu4(NCS)10]^{2-}} copper thiocyanate cluster with antiferromagnetic order [Fig.~\ref{fig:orbital}(b)]. The calculated highest-occupied molecular orbital is consistent with the \ce{N=C=S-} valence-bond resonance form, where the N-terminus is approximately $sp^2$-hybridized. This N $sp^2$ orbital can bond with the Cu $d_{x^2-y^2}$ orbital due to the C--N--Cu bond angle of $165^\circ$. Thus, we can consider the superexchange as occurring through a single molecular orbital where a simultaneous transfer of both spins allows the two half-filled $d_{x^2-y^2}$ orbitals to couple antiferromagnetically [Fig.~\ref{fig:orbital}(c)]. This mechanism is consistent with a correlation-type superexchange, and despite the near-$90^\circ$ Cu--S--C angle, the single-orbital character of the superexchange means the coupling is antiferromagnetic, as predicted by the Goodenough-Kanamori rules.\cite{goodenough_magnetism_1963}

\subsection{Spin density from NMR measurements}
\begin{figure}
	\centering
	\includegraphics[scale=0.5]{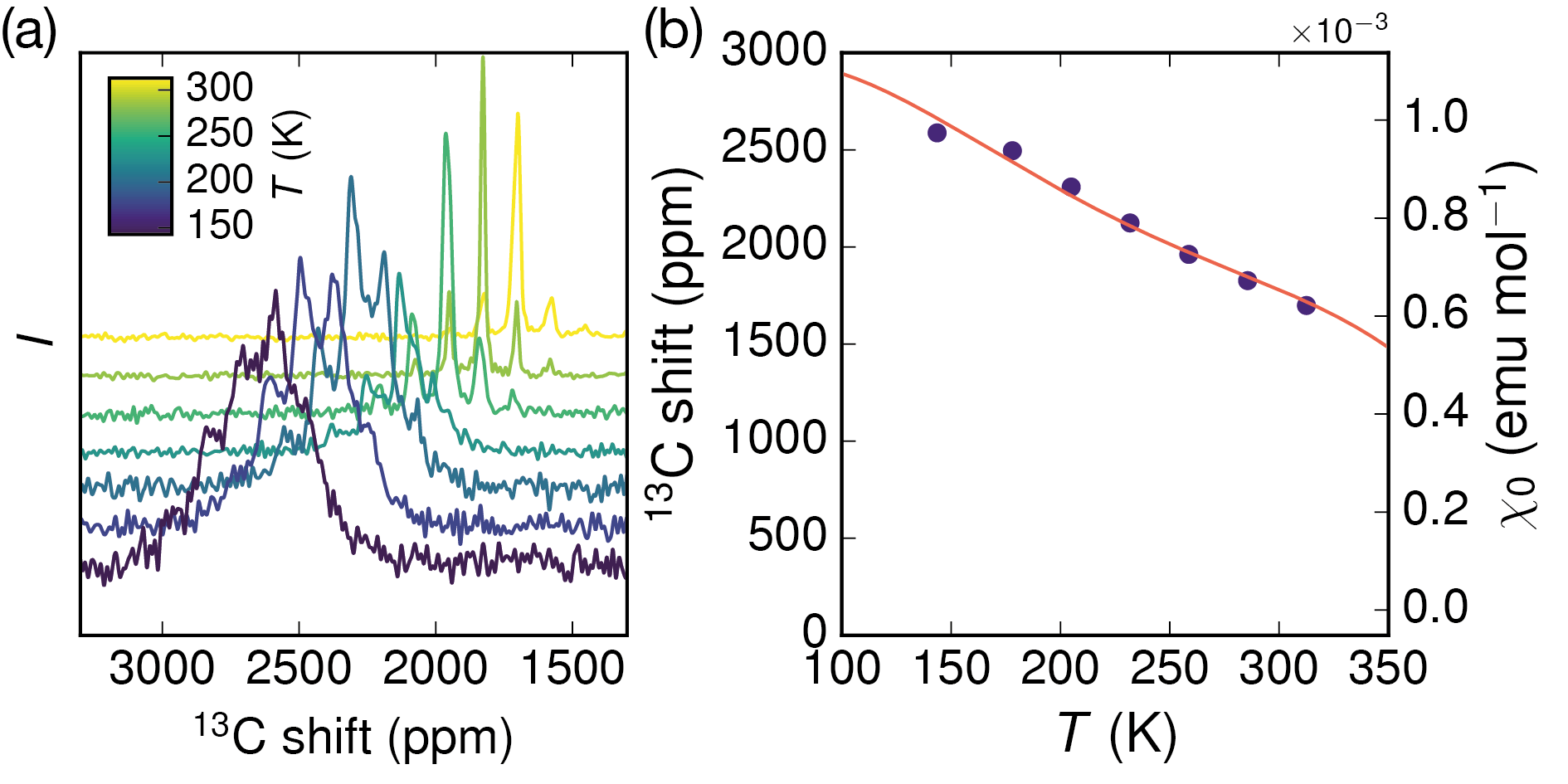}
	\caption{(a) $^{13}$C MASNMR spectra illustrating the evolution of $\delta_\mathrm{iso}$ with temperature. (b) Linear fitting of the experimental isotropic chemical shift, $\delta_\mathrm{iso}$, (left) to the experimental MPMS-measured magnetic susceptibility, $\chi_0$, (right) to extract a value for $\Phi$ (see Eq.~\eqref{eqn:diso}).}
	\label{fig:NMR}
\end{figure}
The high sensitivity of the NMR isotropic chemical shift $\delta\mathrm{_{iso}}$ to the local magnetic susceptibility means that NMR can probe the local spin distribution. Consequently, NMR measurements provide a strong check on the validity of calculated spin and electron densities, in particular through the extent of paramagnetic spin-density transfer. We therefore measured $^{13}$C MASNMR spectra over the range 144--313\,K, to establish the reliability of our calculations. From these spectra we extracted the isotropic $^{13}$C shift $\delta_\mathrm{iso}$ shown in Fig.~\ref{fig:NMR}(a), which can be related to the magnetic susceptibility $\chi_0$ as
\begin{equation}
\delta_\mathrm{iso}=\delta\mathrm{_{iso}^{CS}}+\Phi\chi_0,
\label{eqn:diso}
\end{equation}
where $\delta\mathrm{_{iso}^{CS}}$ is the chemical shift due to the chemical shielding of electrons and $\Phi$ a constant related to the paramagnetic spin-density transfer. In this case, we have taken $\delta\mathrm{_{iso}^{CS}}=133.1$ ppm, as this remains approximately constant in diamagnetic \ce{NCS-} materials.\cite{Dickson1987,Gysling1982,Bowmaker1998} By linear fitting of $\delta_\mathrm{iso}$ to the experimental magnetic susceptibility (measured in an applied field of 0.01\,T) we were able to extract a value for $\Phi$ of $2.52 \times \,10^{6}$\,mol\,emu$^{-1}$ [Fig.~\ref{fig:NMR}(b)]. This compares to the PBE+$U$ value of $1.71\times10^{6}$ mol emu$^{-1}$ [Table \ref{tab:nmr}], demonstrating the reliability of the calculations. Finally, we note that our DFT+$U$ calculations using plane-wave basis sets are also consistent with our hybrid calculations using Gaussian basis sets. These hybrid calculations provided us with the hyperfine parameters for $^{63}$Cu and $^{65}$Cu: principal values for $^{63}$Cu of $A_{AA}=-467.99$\,MHz, $A_{BB}=304.03$\,MHz and $A_{CC}=163.96$\,MHz, and for $^{65}$Cu of $A_{AA}=-500.78$\,MHz, $A_{BB}=325.34$\,MHz and $A_{CC}=175.45$\,MHz. 

\begin{table}
	\centering
	\caption{NMR scaling factor $\Phi$ from DFT and $^{13}$C NMR.}
	\label{tab:nmr}
	\begin{tabular}{ l r }
		& $\Phi$ (mol emu$^{-1}$) \\
		Expt& $2.52\times10^{6}$ \\
		PBE+$U$ & $1.71\times10^{6}$ \\
		Hyb20& $2.00\times10^{6}$ \\
		Hyb35& $1.88\times10^{6}$ \\
	\end{tabular}
\end{table}

\subsection{Spin Hamiltonian from bulk magnetic measurements}
\begin{table}
\centering
\caption{\label{tab:chi_fit} Fitted interaction constants and $g_\mathrm{iso}$.}

\begin{tabular}{ l r r r r r}
& $J_2^\mathrm{1D}$\,(K) & $J_2^\mathrm{2D}$\,(K) & $J_1^\mathrm{2D}$\,(K) &  $g_\mathrm{iso}^\mathrm{1D}$ &  $g_\mathrm{iso}^\mathrm{2D}$  \\\hline
MPMS & $136.5(4)$ &  $133.88(7)$  &  $10.71(6)$ & 2.032(4) & 2.056(1) \\
EPR  & $137(3)$ &  $132.6(1.7) $  &  $17(2)$ & 2.101* & 2.101* \\
\end{tabular}

*$g_\mathrm{iso}$ derived from fitting EPR spectrum at $5\,$K.
\end{table}

\begin{figure}[b]
\centering
 \includegraphics[scale=0.5]{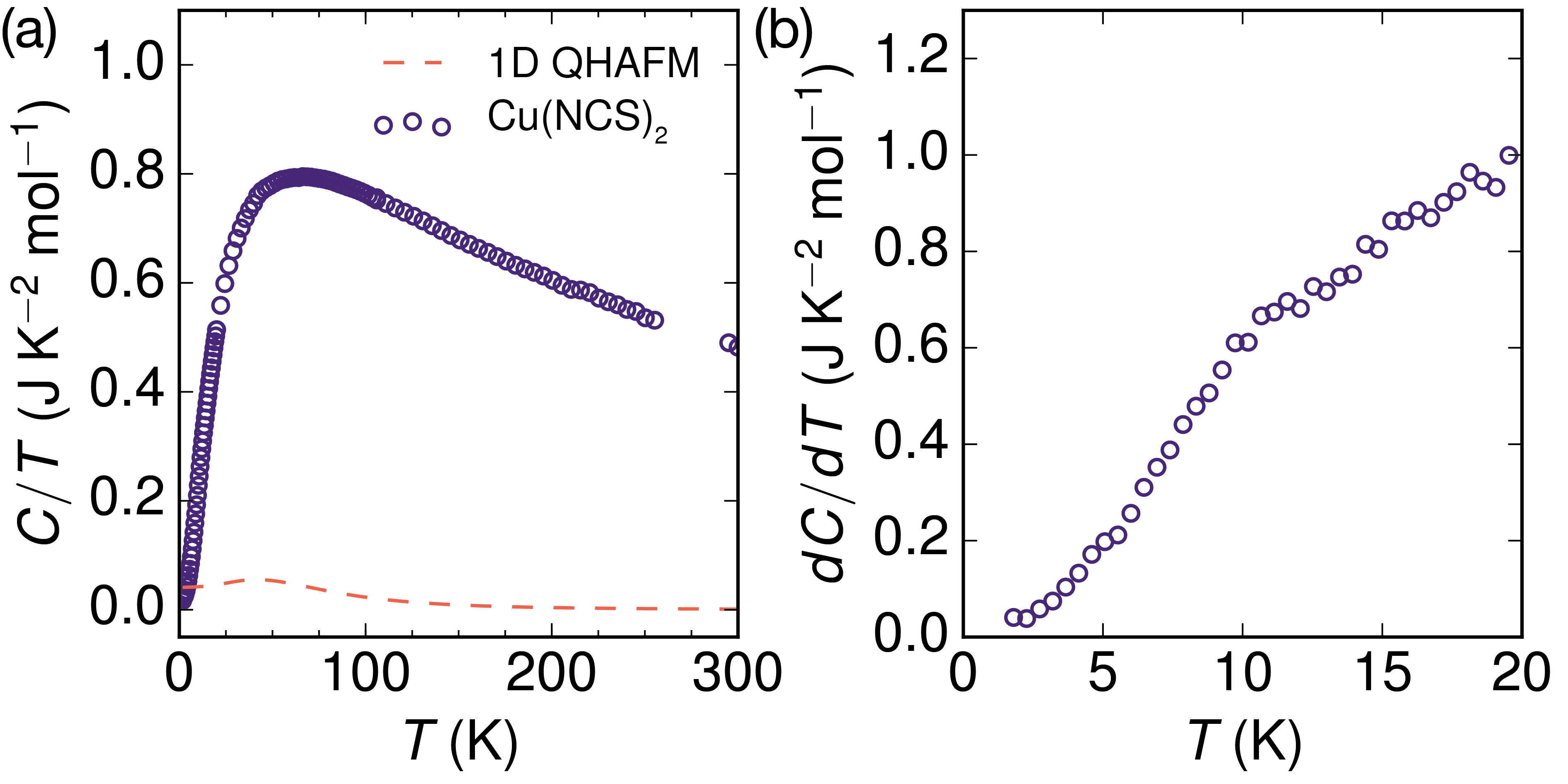}
 \caption{(a) Specific heat divided by temperature, $C$/$T$, for Cu(NCS)$_2$ including both magnetic and lattice contributions. The theoretical heat capacity of a 1D QHAFM is included for comparison. (b) Temperature derivative of the specific heat capacity, d$C$/d$T$, plotted over the low temperature region, to highlight the anomaly at around $T_\mathrm{N}$=12\,K.}
 \label{fig:heatcap}
\end{figure}

As our quantum-chemical calculations suggested that the magnetic interactions were likely to be low dimensional, we carried out fits to the paramagnetic part of both the MPMS and EPR data sets ($T>14\,$K) using low-dimensional models, including both a 1D uniform-chain model (fitting $J_2$ and $g_\mathrm{iso}$, fixing $J_1=J_3=J_4=0$) \cite{Klumper2000} and the 2D coupled chain or rectangular antiferromagnet model (fitting $J_1$, $J_2$ and $g_\mathrm{iso}$, fixing $J_3=J_4=0$).\cite{Keith2011}
We found that the 1D- and 2D-QHAFM models were able to fit both data sets well, but the 2D model was better able to account for the lower-temperature range ($T<30\,$K) [Fig.~\ref{fig:susceptibility}(a,b)]. The independent fits of the MPMS and EPR derived susceptibilities yielded comparable values for the exchange constants, with a large $J_2^\mathrm{2D,MPMS}$=133.88(7)\,K and relatively small $\alpha^\mathrm{2D,MPMS}=J_1^\mathrm{2D,MPMS}/J_2^\mathrm{2D,MPMS}=0.08$ [Table~\ref{tab:chi_fit}]. These fitted values are broadly consistent with those predicted by our DFT calculations, although the calculations overestimate the overall strength of the interactions. The small discrepancy ($\sim$3\%) between the fitted value of $g_\mathrm{iso}$ and that derived from the peak shapes of the EPR spectra is typical for MPMS-derived measurements of $g$, and is likely due to small errors in experimental sample mass. 

The strength of the magnetic interactions here is comparable to the purely inorganic copper(II) halides despite their much-reduced distance between metal centers. However, the lower $T_\mathrm{N}$ of \ce{Cu(NCS)2} means it can be considered a more low-dimensional system; by comparison, for \ce{CuCl2} $T_\mathrm{max}=70$\,K and $T_ \mathrm{N}=23.9$\,K ($l = 3.0$),\cite{Barraclough1964,Banks2009} and for \ce{CuBr2} $T_\mathrm{max}=226$\,K and $T_ \mathrm{N} = 73.5$\,K ($l = 3.1$).\cite{Barraclough1964,Zhao2012} Isothermal magnetisation measurements also showed evidence of a metamagnetic transition at an applied field of 1.4\,T in the ordered phase [Fig.~\ref{fig:susceptibility}(e,f)]. This transition showed no evidence of hysteresis. The low field at which this transition occurs relative to the strong intrachain coupling indicates that it is probably a spin-flop transition, in which the spins rotate away from their easy axis to lie perpendicular to the applied field.\cite{Bogdanov2007} Further analysis was restricted because for our powder sample the applied field will make a random angle with the easy axis, broadening the transition.

Heat-capacity measurements on a pelleted sample of \ce{Cu(NCS)2} did not show a pronounced anomaly at $T_\mathrm{N}$ [Fig.~\ref{fig:heatcap}(a)], consistent with the presence of low-dimensional magnetic correlations above $T_\mathrm{N}$. In this context, we note that for a uniform 1D QHAFM, on cooling to $T_ \mathrm{N}=0.08J$ only 8\% of the magnetic entropy would remain unaccounted for.\cite{Klumper2000} Examination of the temperature derivative of heat capacity showed a change in slope near to $T_ \mathrm{N}$ which may be indicative of the ordering transition [Fig.~\ref{fig:heatcap}(b)].
The small magnitude of the magnetic heat capacity relative to the phononic contribution prevented us from reliably extracting it, as we lacked both a suitable diamagnetic analogue and an appropriate model to account for the anisotropic vibration. A model incorporating two different Debye temperatures (105\,K and 450\,K) was able to account for the majority of the high-temperature heat capacity, reflecting the strongly anisotropic phonons expected for a chain like-structure, but was insufficiently quantitative to extract the magnetic heat capacity.

\subsection{Magnetic ground state from neutron-diffraction measurements}
\begin{figure}
\centering
 \includegraphics[scale=0.51]{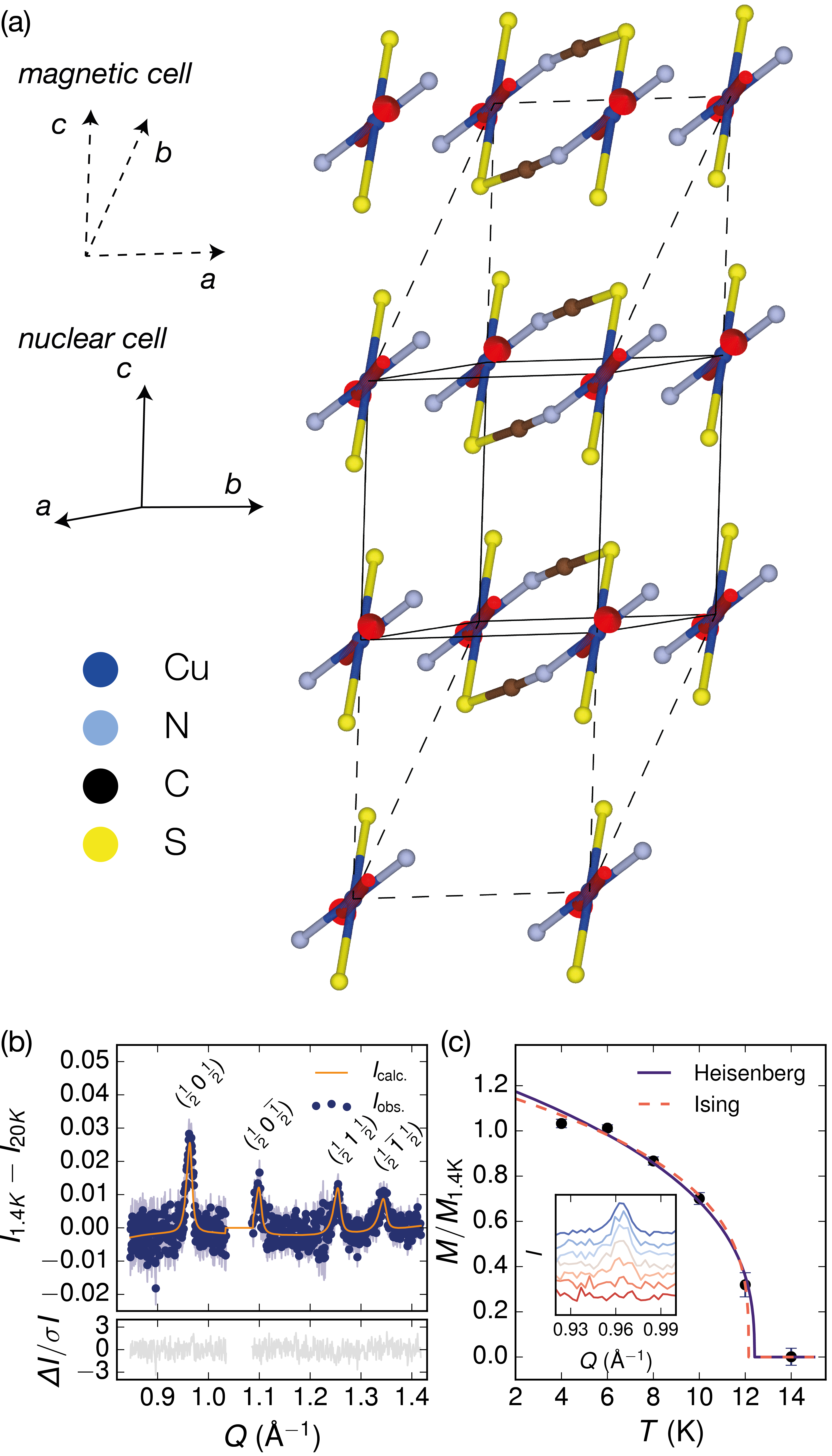}
 \caption{(a) Magnetic structure of \ce{Cu(NCS)2} determined at 2\,K. The nuclear cell is indicated by a solid line, and the magnetic cell by dashed lines. Atoms are colored as follows: Cu dark blue, N light blue, C black and S yellow. The directions of the ordered moments are shown by red arrows. Weak Cu--S interactions are shown as black dashed lines and the \ce{CuN4} coordination polyhedron is indicated by a light blue square. (b) Rietveld refinement of neutron scattering data for \ce{Cu(NCS)2}, with nuclear peaks and background subtracted using a high-temperature data set. The magnetic peaks are labelled with Miller indices corresponding to the nuclear cell. (c) Temperature dependence of the ordered magnetic moment, fitted using the equations for both 3D Heisenberg ($\beta = 0.365$) and Ising magnets ($\beta = 0.302$). The inset shows the temperature evolution of the $(\frac{1}{2} 0 \frac{1}{2})$ reflection on heating from 1.4\,K to 14\,K.}
 \label{fig:mag_structure}
\end{figure}

\begin{table}[h]
\centering
\caption{\label{tab:chi_fit} Components of the ordered magnetic moment derived from Rietveld refinement to neutron-diffraction data measured at 1.4\,K.}

\begin{tabular}{ l l l}
\hline
$M_a$ ($\mu_\mathrm{B}$)&$M_b$ ($\mu_\mathrm{B}$)&$M_c$ ($\mu_\mathrm{B}$)\\
0.311(8)&$-0.413(17)$& 0.278(16) \\
\hline

\end{tabular}

Estimated standard errors in parentheses.
\end{table}

Having established that the paramagnetic phase \ce{Cu(NCS)2} is well-described by a 2D QHAFM Hamiltonian, we were then led to investigate the magnetic ground state. We were motivated in particular because the rapid decrease in magnetic susceptibility below $T_\mathrm{N}$ is consistent with either a spin-Peierls lattice-driven transition to a singlet state or a conventional transition to a long-range ordered antiferromagnet.\cite{Hase1993} We therefore carried out low-temperature neutron-diffraction measurements in order to ascertain the  ground state. On cooling below $T_ \mathrm{N}$ we observed additional superlattice reflections corresponding to a $\mathbf{k}=[\frac{1}{2} 0 \frac{1}{2}]^*$ propagation vector [Fig.~\ref{fig:mag_structure}(a,b)]. The intensities of these weak reflections decreased rapidly with increasing $Q$ and were not accompanied by a significant structural distortion, indicating that this is a long-range ordered antiferromagnetic ground state. 

Symmetry analysis carried out using the ISODISTORT software package showed that only a single irreducible representation, $mU_{1}^{+}$ in Miller and Love's notation,\cite{Cracknell1979} is consistent with the observed propagation vector, which gives a magnetic space group of $P_{s}\overline{1}$. The resulting magnetic unit cell is related to the nuclear unit cell by the transformation matrix
$$
\begin{pmatrix} 
\mathbf{a}' \\ 
\mathbf{b'} \\
\mathbf{c}' \\
\end{pmatrix}
=
\begin{pmatrix} 
0  & 1 & 0\\
-1 & 0 & 1\\
0  & 0 & 2\\
\end{pmatrix}
\begin{pmatrix} 
\mathbf{a}\\ 
\mathbf{b}\\
\mathbf{c}\\
\end{pmatrix}.
$$
The ordered structure is shown in Fig.~\ref{fig:mag_structure}(a), and is consistent with the calculated exchange constants and can be considered as a $G$-type ordering on the cuboidal lattice formed by the three shortest non-coplanar Cu--Cu distances ([100], [101] and [001]). Rietveld refinement of the magnetic structure against the 1.4\,K data set (with nuclear scattering removed by subtraction of a high-temperature data set) gave a nearly saturated ordered moment $M_0=0.30(3)\,\mu_\mathrm{B}$ per Cu [Fig.~\ref{fig:mag_structure}(b)]. This strong reduction in the ordered moment, compared to the maximum value of $1\,\mu_\mathrm{B}$ per Cu, is consistent with the expected highly-fluctuating nature of low-dimensional quantum magnetism even in the ordered phase. It is also comparable to the mean-field estimate for coupled 1D QHAFM chains, $M_0/\mu_\mathrm{B} \approx \frac{\alpha}{2}^{\frac{1}{2}} \approx 0.2$, where the interchain coupling has been approximated as $(J_1+J_4)/2$.\cite{Kojima1997,Schulz1996,Ge2017} Refinement of the moment direction revealed that it lies within the plane of the Cu $d_{x^2-y^2}$ orbital. The temperature dependence of the magnetic order parameter could be fitted to $M/M_\mathrm{1.4\,K} = A (T_ \mathrm{N}-T)^\beta$, $\beta=0.33(3)$, which is consistent with this transition belonging to a 3D universality class [Fig.~\ref{fig:mag_structure}(c)]; however we were unable to distinguish between different spin anisotropies due to the relative weakness of the magnetic reflections.\cite{Campostrini2002} 
\section{Conclusions}
Our study reports the magnetic structure, single-ion properties and magnetic interactions of a thiocyanate-based molecular framework, and resolves the long-standing question of the chemical identity and structure of this simple binary pseudohalide.\cite{Claus1838} We have also demonstrated that \ce{Cu(NCS)2} is a low-dimensional magnet, which behaves at high temperatures as a 1D QHAFM and at intermediate temperatures as a rectangular 2D QHAFM, before ordering as a three-dimensional antiferromagnet. The significantly reduced ordered moment suggests that, like \ce{LiCuVO4},\cite{Naito2007} \ce{Li2CuZrO4}\cite{Drechsler2007} and \ce{Cu(DCO2)2.4D2O},\cite{DallaPiazza2014} signatures of low dimensional quantum behavior persist even in the 3D ordered state.

The categorisation of \ce{NCS-} as a pseudohalide---that is, a molecular anion that behaves like a single atom halide anion---was a concept developed to explain the structures and reactivities of \ce{NCS-} based materials.\cite{Birckenbach1925} Our results give further weight to the use of this analogy for rationalizing magnetic properties.

Nevertheless, careful design will be necessary to exploit these ligands fully. In the related materials \ce{Cu(NCS)2(pyrazine)2}\cite{Bordallo2003} and 
\ce{Cu(NCS)2(pyrimidine)2},\cite{Wriedt2009} chemical substitutions that might be expected to reduce the dimensionality of the magnetic interactions by suppressing interactions between the \ce{Cu(NCS)2} chains instead reorient the magnetic $d_{x^2-y^2}$ orbital out of the plane containing the chain, reducing the superexchange interaction from $J=133$\,K to $J=-1.00$\,K.\cite{Wriedt2009} In molecular frameworks, therefore, the sensitivity of the overlap integral to ligand orientation can be increased compared to inorganic materials. The heightened propensity of molecular framework materials to undergo large structural distortions thus also suggests that they will prove valuable systems to study coupling between mechanical and magnetic degrees of freedom.\cite{Brinzari2013,Quintero2015}

\acknowledgements
M.J.C. thanks Sidney Sussex College, University of Cambridge for financial support. J.A.M.P. thanks Churchill College, University of Cambridge for financial support. J.L. thanks Trinity College Cambridge for financial support. S.S. and P.M. acknowledge funding from the Winton Programme for the Physics of Sustainability. M. W.G. thanks the European Union's Horizon 2020 research and innovation programme for support under the Marie Sk{\l}odowska--Curie grant agreement No. 659764. Computational resource was provided by the Center for Functional Nanomaterials, Brookhaven National Laboratory, which is supported by the U.S. Department of Energy, Office of Basic Energy Sciences, under Contract No. DE-AC02-98CH10886. We acknowledge the Rutherford Appleton Laboratory for access to the ISIS Neutron Source. Use of the Advanced Photon Source at Argonne National Laboratory was supported by the U. S. Department of Energy, Office of Science, Office of Basic Energy Sciences, under Contract No. DE-AC02-06CH11357. Magnetic measurements were carried out using the Advanced Materials Characterisation Suite, funded by EPSRC Strategic Equipment Grant EP/M000524/1. We thank Evan N. Keyzer for his assistance with diffuse reflectance measurements and Manoranjan Kumar and Andrew J. Pell for helpful discussions.

\bibliography{CuSCN2}

\end{document}